\documentclass[preprint2]{emulateapj}

\newcommand{\bzcat}{ROMA-BZCAT}

\newcommand{\fer}{{\it Fermi}}

\usepackage{graphicx}
\usepackage{longtable}

\slugcomment{version \today: fm}
\shorttitle{The low-frequency radio catalog of flat spectrum sources}
\shortauthors{F. Massaro et al. 2014}

\begin{document}
\title{The low-frequency radio catalog of flat spectrum sources}
\author{
F. Massaro\altaffilmark{1,2}, 
M. Giroletti\altaffilmark{3},
R. D'Abrusco\altaffilmark{4}, 
N. Masetti\altaffilmark{5},
A. Paggi\altaffilmark{4}, \\
Philip S. Cowperthwaite\altaffilmark{4},
G. Tosti\altaffilmark{6}
\& 
S. Funk\altaffilmark{2}.
}

\altaffiltext{1}{SLAC National Laboratory and Kavli Institute for Particle Astrophysics and Cosmology, 2575 Sand Hill Road, Menlo Park, CA 94025, USA}
\altaffiltext{2}{Yale Center for Astronomy and Astrophysics, Physics Department, Yale University, PO Box 208120, New Haven, CT 06520-8120, USA}
\altaffiltext{3}{INAF Istituto di Radioastronomia, via Gobetti 101, 40129, Bologna, Italy}
\altaffiltext{4}{Harvard - Smithsonian Astrophysical Observatory, 60 Garden Street, Cambridge, MA 02138, USA}
\altaffiltext{5}{INAF - Istituto di Astrofisica Spaziale e Fisica Cosmica di Bologna, via Gobetti 101, 40129, Bologna, Italy}
\altaffiltext{6}{Dipartimento di Fisica, Universit\`a degli Studi di Perugia, 06123 Perugia, Italy}

\begin{abstract}
A well known property of the $\gamma$-ray sources detected by COS-B in the 1970s, by the Compton Gamma-ray Observatory in the 1990s
and recently by the \fer\ observations is the presence of radio counterparts, in particular for those associated to extragalactic objects. 
This observational evidence is the basis of the radio-$\gamma$-ray connection established for the class of active galactic nuclei known as blazars.
In particular, the main spectral property of the radio counterparts associated with $\gamma$-ray blazars is that they show a flat spectrum in the GHz
frequency range. Our recent analysis dedicated to search blazar-like candidates as potential 
counterparts for the unidentified $\gamma$-ray sources (UGSs) allowed us to extend the radio-$\gamma$-ray connection in the MHz regime.
We also showed that below 1 GHz blazars maintain flat radio spectra. Thus on the basis of these new results, we assembled a 
low-frequency radio catalog of flat spectrum sources built by combining the radio observations of the Westerbork Northern Sky Survey (WENSS)
and of the Westerbork in the southern hemisphere (WISH) catalog with those of the NRAO Very Large Array Sky survey (NVSS).
This could be used in the future to search for new, unknown blazar-like counterparts of the $\gamma$-ray sources.
First we found NVSS counterparts of WSRT radio sources and then we selected flat spectrum radio
sources according to a new spectral criterion specifically defined for radio observations performed below 1 GHz.
We also described the main properties of the catalog listing 28358 radio sources and their logN-logS distributions. 
Finally a comparison with with the Green Bank 6-cm radio source catalog has been performed 
to investigate the spectral shape of the low-frequency flat spectrum radio sources at higher frequencies.
\end{abstract}


\section{Introduction}
\label{sec:intro}
Since the epoch of the first $\gamma$-ray observations performed by COS-B 
in the 1970s \citep[e.g.,][]{hermsen77} and by the 
Compton Gamma-ray Observatory in the 1990s \citep[e.g.,][]{hartman99}, 
a link between the radio and the $\gamma$-ray sky was discovered. It has been used to 
associate the high-energy sources with their low-energy counterparts \citep[e.g.][]{mattox97}.
This radio-to-$\gamma$-ray relation has also been recently highlighted for the extragalactic sources 
detected by the \fer\ mission \citep{atwood09}.
In particular, nearly all the $\gamma$-ray sources associated in the second \fer\ 
Large Area Telescope (LAT) catalog \citep[2FGL;][]{nolan12} 
and/or in the second catalog of active galactic nuclei (AGNs) \citep{ackermann11a} 
detected by the \fer-LAT have a clear radio counterpart. 
This is the basis of the radio-$\gamma$-ray connection specifically discussed for 
blazars \citep[e.g.,][]{ghirlanda10,mahony10,ackermann11b},
that constitute the rarest class of AGNs \citep[e.g.,][]{urry95,massaro09,massaro11} and the largest known 
population of $\gamma$-ray sources \citep[e.g.,][]{abdo10}.

Recently we addressed the problem of searching for $\gamma$-ray blazar candidates 
as counterparts of the unidentified $\gamma$-ray sources (UGSs) adopting a new approach that
employs the low-frequency radio observations performed by the Westerbork Synthesis Radio Telescope (WSRT). 
While performing this investigation we found that the radio-$\gamma$-ray connection of blazars 
can be extended below $\sim$1 GHz \citep{ugs3}.

Our analysis was based on the combination of the 
radio observations from Westerbork Northern Sky Survey \citep[WENSS;][]{rengelink97}
at 325 MHz with those of the NRAO Very Large Array Sky survey
\citep[NVSS;][]{condon98} and of the Very Large Array Faint Images of the Radio Sky 
at Twenty-Centimeters \citep[FIRST;][]{becker95,white97} at about 1.4 GHz.
A similar analysis was also performed using the Westerbork in the 
southern hemisphere (WISH) survey \citep{debreuck02} at 352 MHz \citep{ugs6}.
Both of these studies were based on the observational evidence that blazars also show flat radio spectra below $\sim$1 GHz 
\citep[see also][for recent analyses]{kovalev09a,kovalev09b,petrov13}.

The flatness of the blazar radio spectra is a well known property expected from   
radio data in the GHz frequency range \citep[e.g.,][for recent analyses]{ivezic02,healey07,kimball08}.
This spectral property was also used in the past for the associations of $\gamma$-ray sources
since the EGRET era \citep[e.g.,][]{mattox97}.
However, despite a small survey of BL Lac objects at 102 MHz \citep{artyukh81},
the low radio frequency spectral behavior of blazars was still an unexplored region of the electromagnetic spectrum
until our recent analyses \citep{ugs3,ugs6}.
Using WSRT data at 325 MHz and at 352 MHz as well as those of very low-frequency observations of the 
Very Large Array Low-Frequency Sky Survey\footnote{http://lwa.nrl.navy.mil/VLSS/} \citep[VLSS;][]{cohen07}
at 74 MHz we showed that blazars maintain a flat radio spectrum even below $\sim$100 MHz and we extended the 
radio-$\gamma$-ray connection below $\sim$1 GHz \citep{massaro13b}.

Thus, motivated by these recent results we assembled a catalog of low-frequency flat spectrum radio sources
using the combination of both the WENSS and the WISH surveys with the NVSS.
The main aim of this investigation is to provide the counterpart, at longer wavelengths, of the 
Combined Radio All-Sky Targeted Eight-GHz Survey (CRATES) used to associate
\fer\ objects with blazar-like sources \citep{healey07}.

The paper is organized as follows: in Section~\ref{sec:survey} we briefly present the main properties of the 
low-frequency radio survey performed by WSRT and used to carry out our investigation (i.e., the WENSS and the WISH).
In Section~\ref{sec:radius} we search for the NVSS counterparts of WSRT sources.
Then in Section~\ref{sec:catalog} we extract the main 
low-frequency catalog of flat spectrum radio sources (LORCAT) 
from the combined WSRT-NVSS surveys and we discuss on its main properties. 
Section~\ref{sec:cross} is devoted to the comparison the Green Bank 6-cm (GB6) 
radio source catalog \citep[e.g., ][]{gregory96} to investigate the spectral behavior of LORCAT sources at higher frequencies.
Finally, Section~\ref{sec:conclusions} is dedicated to the summary and the conclusions.

For our numerical results, we use cgs units unless stated otherwise
Spectral indices, $\alpha$, are defined by flux density, S$_{\nu}\propto\nu^{-\alpha}$.
The WSRT catalogs used to carry out our analysis are available from both the 
HEASARC\footnote{WENSS:http://heasarc.gsfc.nasa.gov/W3Browse/all/wenss.html}$^,$\footnote{WISH: http://heasarc.gsfc.nasa.gov/W3Browse/all/wish.html} 
and the VIZIER\footnote{WENSS: http://vizier.u-strasbg.fr/viz-bin/VizieR?-source=VIII/62}$^,$\footnote{WISH: http://cdsarc.u-strasbg.fr/viz-bin/Cat?VIII/69A} 
databases as well as that of the NVSS\footnote{http://heasarc.gsfc.nasa.gov/W3Browse/all/nvss.html}$^,$\footnote{http://vizier.u-strasbg.fr/viz-bin/VizieR?-source=\%20NVSS}.

\section{Westerbork Low-frequency radio survey}
\label{sec:survey}
The Westerbork Northern Sky Survey (WENSS) is a low-frequency radio survey that covers the northern sky above +30\degr\ in declination 
performed at 325 MHz to a limiting flux density of $\sim$18 mJy at 5 sigma level \citep{rengelink97}. 
The version of the WENSS catalog used in our analysis was implemented as a combination 
of two separate catalogs obtained from the WENSS Website\footnote{http://www.astron.nl/wow/testcode.php?survey=1}: 
the WENSS Polar Catalog that comprises 18186 sources above +72\degr\ in declination 
and the WENSS Main Catalog including 211234 objects in the declination range between +28\degr\ and +76\degr.

We also used the Westerbork In the Southern Hemisphere (WISH) catalog\footnote{http://www.astron.nl/wow/testcode.php?survey=2} 
that is the southern extension of the WENSS. WISH is a low-frequency (352 MHz) radio survey 
covering most of the sky between -26\degr and -9\degr\ at 352  MHz to the same limiting flux density of the WENSS. 
It is worth noticing that the Galactic Plane region at galactic latitudes $|b|<$10\degr\ are excluded from the WISH observations.
Due to the very low elevation of the observations, the survey has a much lower resolution in declination than in right ascension. 
A correlation with the NVSS shows that the positional accuracy is less constrained in declination than in right ascension, 
but there is no significant systematic error \citep[see][for more details]{debreuck02}. 
Finally, we highlight that the WISH catalog contains multiple observations of the same source 
for many objects as well as measurements of individual components of multi-component sources.

\section{Radio spatial associations}
\label{sec:radius}
We adopted the following statistical approach to find the radio NVSS counterparts at 1.4 GHz for the sources 
in the WSRT low radio frequency surveys, namely: the WENSS and the WISH.

For each radio source listed in either the WENSS and the WISH surveys, we searched for all the NVSS counterparts 
that lie within elliptical regions that corresponds to the positional uncertainty at 95\% level of confidence (i.e., 2$\sigma$).
We took into account the uncertainties on both the right ascension, $\alpha$, and the declination, $\delta$, in the WSRT and in the NVSS surveys.
 
We found that the total number of correspondences is 225933 out of 268425 radio sources included in either the WSRT surveys.
We excluded from our analysis all the WSRT source with radio analysis flags 
{ (i.e., P and Y as reported in the WENSS and WISH catalog, respectively,
to indicate that there were problems in the model fitting for a source)}
and variability flag in the WISH observations, all the double matches and all those sources labeled 
as components of a multi-component source (flag ``C") in the WSRT catalogs.
In addition, for this version of the LORCAT catalog, 
we also excluded from our sample 2707 multiple matches since their WSRT radio flux densities
could be due to the emission of several, unresolved, NVSS sources 
so contaminating our estimates of the low frequency spectral index.

We then built 100 mock realizations of the WSRT catalog by shifting each source position
in a random direction of the sky by a fixed length of 1\degr.
This shift adopted to create the mock WSRT catalogs were chosen not too distant 
from the original WSRT position and within the NVSS footprint
so to obtain fake catalogs with a sky distribution similar to the original WSRT
and to perform the cross-match with each fake catalog and the NVSS taking 
into account the local density distribution of the WSRT radio sources.
The total number of WSRT sources in each mock realization is also preserved.

For each mock realization of the WSRT catalog, we counted the number of associations with the NVSS occurring 
at angular separations $R$ smaller than 300\arcsec. Then we computed the mean number $\lambda(R)$ of these 
mock associations, averaged over the 100 fake WSRT catalogs and verifying that $\lambda(R)$ has a Poissonian distribution.
Increasing the radius by $\Delta\,R=$5\arcsec, we also computed the difference $\Delta\,\lambda(R)$ as:
\begin{equation}
\Delta\,\lambda(R) = \lambda(R+\Delta\,{R}) - \lambda(R)\,,
\end{equation}

\begin{figure}[] 
          \includegraphics[height=9.5cm,width=7.2cm,angle=-90]{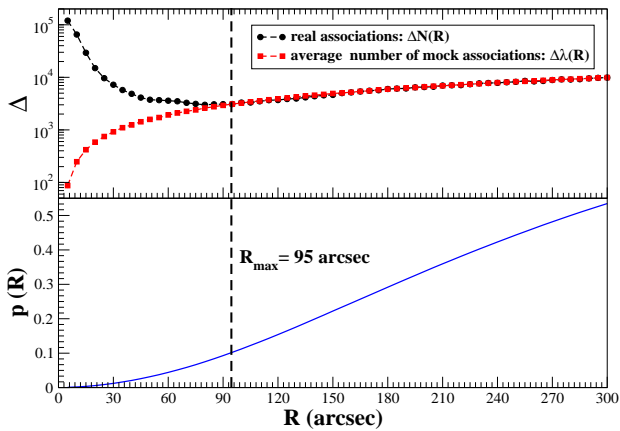}
           \caption{Upper panel) The values of $\Delta\,\lambda(R)$ (red circles) and $\Delta\,N(R)$ (black squares) 
                        as function of the angular separation $R$.
                        Our choice of $R_{max}$ is marked by the vertical dashed line. It occurs at the first 
                        $R$ value for which $\Delta\,\lambda(R)\simeq\Delta\,N(R)$.
                        Lower panel) The probability of having spurious associations $p(R)$
                        as function of the angular separation $R$.
                        }
          \label{fig:delta}
          \end{figure}

In Figure~\ref{fig:delta} we show the comparison between $\Delta\,N(R)$ and $\Delta\,\lambda(R)$.
For radii larger than $R_{max}=$95\arcsec the $\Delta\,\lambda(R)$ curve superimposes that of $\Delta\,N(R)$
indicating that WSRT-NVSS cross-matches could occur by chance at angular separations larger than $R_{max}$.
Thus we choose $R_{max}$ as to the maximum angular separation between the WSRT and the NVSS position
to consider the 1.4GHz radio source a reliable counterpart of a WSRT object.

In addition we calculated the chance probability of spurious associations $p(R)$ as the ratio 
between the number of real associations $N(R)$
and the average of those found in the mock realizations of the WSRT catalog $\lambda(R)$, corresponding to a value of $\sim$10\%
for $R=R_{max}$ \citep[see e.g.,][for a similar procedure to estimate the probability of spurious associations]{maselli10,paper1,ugs1,ugs2}.

We then computed the uncertainties on the WSRT positions according to the procedure described in Rengelink et al. (1997)
and combined them with the NVSS ones \citep{condon98} using the following relation:
         \begin{equation}
         \sigma_{RA,Dec} = \sqrt{\sigma_{RA,Dec}^2(WSRT)+\sigma_{RA,Dec}^2(NVSS)},
          \end{equation} 
{ We also defined the angular separation normalized to the values of the positional uncertainties $m$ as:
          \begin{equation}
         m = \sqrt{\left(\frac{R_{RA}}{\sigma_{RA}}\right)^2+\left(\frac{R_{Dec}}{\sigma_{Dec}}\right)^2}\\
          \end{equation} 
where $R_{RA}$ and $R_{DEC}$ are the angular separations in right ascension and in declination, respectively.
In Figure~\ref{fig:posunchist} we show the distributions of the ratio between the angular separation $R$ and the 
combined positional uncertainty $\sigma$ for RA and Dec, respectively.
         \begin{figure}[] 
          \includegraphics[height=9.5cm,width=6.2cm,angle=-90]{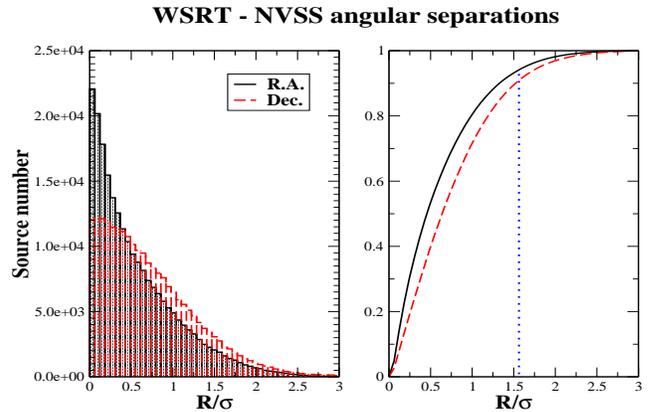}
           \caption{Left panel) The distributions of ratio between angular separations $R$ and the positional uncertainties 
                        for the right ascension (black straight line) and for the declination (red dashed line), respectively for the selected 
                        224438 WSRT - NVSS radio sources.
                        Right panel) The cumulative distribution of the ratio $R/\sigma$. 
                        The dotted blue line marks the 90\% limit (see also Section~\ref{sec:radius}).}
          \label{fig:posunchist}
          \end{figure}
In order to build the final sample of WSRT - NVSS correspondences that will be
used to extract the low-frequency radio catalog of flat spectrum sources, we selected only sources with $m<3$.
This WSRT - NVSS final sample lists 224438 radio sources out of 225933 previously selected. 
We note that we found a potential NVSS counterpart for about 85\% of the WSRT sources,
and since we adopted a threshold on $m=$3, this corresponds to a completeness $C$ of about 80\%,
evaluated according to the relations described in Condon et al. (1975).}
Moreover, this is also in agreement with the reliability of our associations,
estimated via Monte Carlo simulations, occurring at $R_{max}$ that is of the order to 10\% (see Figure~\ref{fig:delta}).

\section{Low-frequency radio catalog of flat spectrum sources}
\label{sec:catalog}

\subsection{Radio spectral index distribution at low-frequencies}
\label{sec:index}
For the WSRT-NVSS associations we defined a low-frequency radio spectral index: $\alpha_{low}$, 
using the integrated flux densities at 325 MHz from the WENSS and those at 352 MHz 
reported in the WISH, $S_{325}$ and $S_{352}$, respectively,
in combination with the NVSS $S_{1400}$ at 1.4 GHz as:
\begin{equation}
\alpha_{low} = - k_1 \cdot log (S_{1400}/S_{low})~,
\end{equation}
{ where the $k_1$ factor is equal to 1.58 and 1.67  (i.e., $[log(1400/325)]^{-1}$ and $[log(1400/352)]^{-1}$) 
for the WENSS and the WISH surveys, respectively, $S_{low}$ is the flux density at 325 MHz (WENSS) or at 352 MHz (WISH)
with all flux densities in units of mJy.}
The uncertainties on $\alpha_{low}$ where computed according to the following relation:
\begin{equation}
\sigma_{low} = k_2 \cdot \sqrt{(\sigma_{1400} / S_{1400})^2 + (\sigma_{low} / S_{low})^2} 
\end{equation}
where the $k_2$ factor is equal to 0.68 and 0.72  (i.e., $\mid [ln(1400/325)]^{-1} \mid$ and $\mid [ln(1400/352)]^{-1} \mid$) 
for the WENSS and the WISH surveys, respectively while $\sigma_{1400}$ and $\sigma_{low}$ 
are the uncertainties on the WSRT and NVSS flux densities.

In radio astronomy it is conventional to indicate flat spectrum radio sources as those 
with a two-point spectral index $\alpha(\nu_1,\nu_2)\sim$0 or typically lower than 0.5 \citep[e.g.][]{condon84a}.
The origin of these thresholds reside in the distribution of the two-point spectral indices measured between 
$\sim$1.4 GHz and $\sim$5 GHz for a number of flux-limited source samples \citep{witzel79,owen83,condon84a,condon88}. 
As shown in these analyses, the (unnormalized) spectral-index distributions consist
of a narrow steep-spectrum component with $\alpha(\nu_1,\nu_2)\sim$0.7 
and a broader flat-spectrum component centered on $\alpha(\nu_1,\nu_2)\sim$0. 
As the sample selection frequency is lowered, the number of steep-spectrum sources increases 
rapidly and the median spectral indices of both components increase \citep[e.g.,][]{kellermann74,condon88}. 
As reported by Kellermann et  al. (1964), the increase in $\alpha(\nu_1,\nu_2)$ 
of each spectral component is proportional to the square of its width,
so the median spectral index of the flat-spectrum component changes more rapidly with frequency. 

\begin{figure}[]  
           \includegraphics[height=9.5cm,width=8.4cm,angle=-90]{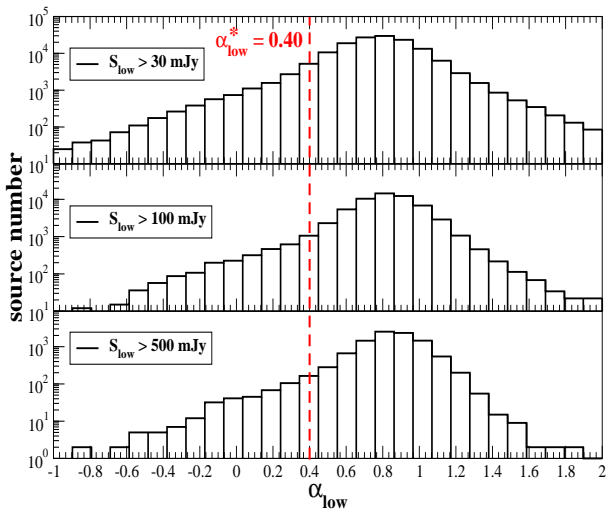}
           \caption{The distributions of $\alpha_{low}$ for three flux limited subsamples 
                         extracted from the WSRT-NVSS correspondences.
                         Flux density cuts are reported in each panel.
                         It is clear how the distribution has a tail for low values of the 
                         low-frequency spectral index but it does not appear to have a bimodal shape
                         as in previous radio analyses at higher frequencies 
                         \citep[see e.g., ][]{witzel79,owen83,condon84a,condon88}. 
                         The red dashed line marks our threshold to define 
                         low-frequency flat spectrum radio sources at $\alpha_{low}$=0.4.}
          \label{fig:distalfas}
          \end{figure}

As shown in Figure~\ref{fig:distalfas}, even considering three or more 
flux limited subsample of the WSRT-NVSS associated sources,
we were not able to identify a bimodal behavior in the spectral index distribution of $\alpha_{low}$.
This, in addition to the frequency dependence highlighted by Kellerman (1964),
suggest that a different criterion has to be chosen to indicate flat spectrum radio sources at low frequencies.

We noticed that both blazars and \fer\ blazars detected in the WENSS show values of $\alpha_{low}$ between -1 and 0.65
for the largest fraction of their samples \citep{ugs3}. Specifically, in our previous analysis we  
considered low-frequency flat spectrum radio sources as those having $\alpha_{low}<$0.65.
This occurred for 90\% of the blazars detected by \fer\
and for more than 80\% of those listed in the \bzcat\ \citep{massaro09,massaro11}.
However, to assemble the LORCAT, we adopted a more conservative threshold based on
the following statistical criterion.

\begin{figure}[] 
          \includegraphics[height=9.5cm,width=7.4cm,angle=-90]{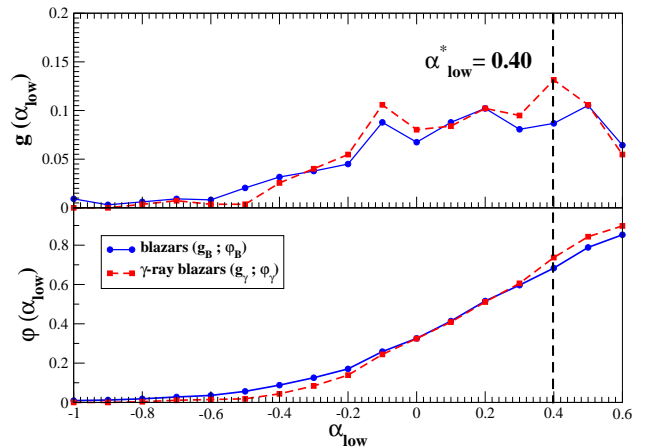}
           \caption{Upper panel) The fractional efficiency $g(\alpha_{low})$ defined by Eq.~\ref{eq:frac}
                          for blazars (blue line), the subsample of \fer\ blazars (red line) 
                          all with $\alpha_{low}$ between -1 and 0.7 (dashed black line).
                          Lower panel) The completeness $\phi$ defined as the ratio between the total number of sources with 
                          ($\alpha_{low}<\alpha^*_{low}$) and the total number of expected sources, computed for blazars (blue line),
                         the \fer\ blazars (red line; see Section~\ref{sec:index} for more details).
                         The chosen threshold $\alpha^*_{low}=$0.4, adopted to create the LORCAT, 
                         is highlighted by the dashed vertical line in both panels.}
          \label{fig:threshold}
          \end{figure}

First we established the number of blazars and \fer\ blazars that we expect to find 
within this subsample simply performing the crossmatch with the \bzcat\ within 8\arcsec.5 as adopted in our previous analysis.
We found that the number of expected blazars with a WSRT counterpart is 979, including 274 \fer\ blazars.
For a given value of the threshold $\alpha^*_{low}$, we defined the fractional efficiency $g(\alpha_{low})$
as the ratio between the difference of total number of sources having: 
$\alpha_{low}<\alpha^*_{low}$ and those with $(\alpha_{low}-\Delta\,\alpha)<\alpha^*_{low}$ 
and the total number of expected sources $N_{exp}$
within the WSRT-NVSS associations with -1$<\alpha_{low}<$0.7:
          \begin{equation}
          g(\alpha_{low}) = \frac{N(\alpha_{low}<\alpha^*_{low})-N((\alpha_{low}-\Delta\,\alpha)<\alpha^*_{low})}{N_{exp}}
          \label{eq:frac}
          \end{equation} 
where $\Delta\,\alpha$= 0.1.
In particular, $g(\alpha_{low})$ has been computed for all blazars (i.e., $g_B(\alpha_{low})$), 
the subsample of \fer\ blazars (i.e., $g_\gamma(\alpha_{low})$)
with -1$<\alpha_{low}<$0.7 (see Figure~\ref{fig:threshold}).
Since the main goal underlying the LORCAT is to have a catalog
of potential counterparts for the UGSs, we chose the $\alpha^*_{low}=$0.4 
threshold as the value corresponding to the peak of the $g_\gamma(\alpha_{low})$.
According to the above threshold the total number of low-frequency sources with a flat radio spectrum 
listed in the LORCAT is 28358 having -1$<\alpha_{low}<$0.40. 
{ Adopting the above criterion on the choice of the $\alpha_{low} $threshold
the LORCAT catalog will be less complete. However, the selected low frequency flat spectrum radio sources 
are more reliable to be $\gamma$-ray blazar candidates since this criterion ensures to avoid
the heavy contamination by steep spectrum radio sources.}

In Figure~\ref{fig:threshold} we also show the completeness $\varphi$ of the sample considered above defined 
as the ratio between the total number of sources with 
Our criterion is then supported by the comparison at high-frequency described in Section~\ref{sec:cross}.
($\alpha_{low}<\alpha^*_{low}$) and the total number of expected sources:
          \begin{equation}
          \varphi(\alpha_{low}) = \frac{N(\alpha_{low}<\alpha^*_{low})}{N_{exp}}
          \label{eq:frac}
          \end{equation} 
Thus we noticed that for our choice of $\alpha^*_{low}=$0.4 we are able to re-associate 80\% of the \fer\ blazars 
of all blazars listed in the WSRT-NVSS with $\alpha_{low}$ between -1 and 0.65 \citep{ugs3}. 

{ In Table~\ref{tab:LORCAT} we listed all LORCAT sources with their WSRT and NVSS names.
For all these\ sources we also report the NVSS coordinates,
the angular separation $R$ between the NVSS and the WSRT positions, the
$\alpha_{low}$ value with its uncertainty $\sigma_{low}$ and the WSRT survey name
which each original source belong to: WISH or WENSS.}

Finally, we note that, as found by our previous analysis \citep[e.g.,][]{ugs3},
the source density of the LORCAT sources is $\sim$1.8 src/deg$^2$, given the total 4.7 sr 
of footprint of the combined WENSS-WISH survey
(3.1 sr in the WENSS  plus 1.6 sr in the WISH) 
while according to the \bzcat\ the blazar density is currently of the order of 0.1 src/deg$^2$.
So we can expect that only about 10\% of the sources in the LORCAT are blazar-like, however, 
to confirm this insight optical spectroscopic observations and high frequency radio data are necessary. 
{ Moreover,} since the \bzcat\ is not a survey and it is not a complete catalog,
the above estimate on the expected fraction of blazars present in the LORCAT has to be considered carefully.
\begin{table*}
\caption{LORCAT Main Table (first 10 lines).}
\tiny
\begin{tabular}{|llllcccl|}
\hline
  WSRT  & NVSS &  R.A. (NVSS)   & Dec. (NVSS)  & R            & $\alpha_{low}$ & $\sigma_{low}$  & survey \\
  name   & name   &  (J2000)             & (J2000)           & arcsec   &                             &                              &              \\
\hline
\noalign{\smallskip}
  0000.0+3323 & J000238+334008 & 00:02:38.41 & +33:40:08.3 & 6.27 & 0.18 & 0.12 & wenss \\
  0000.0+4449 & J000237+450554 & 00:02:37.65 & +45:05:54.1 & 10.98 & 0.38 & 0.1 & wenss \\
  0000.0+5005 & J000236+502220 & 00:02:36.82 & +50:22:20.3 & 6.78 & -0.28 & 0.12 & wenss \\
  0000.0+6737 & J000235+675422 & 00:02:35.79 & +67:54:22.7 & 2.78 & 0.33 & 0.07 & wenss \\
  0000.0-1838 & J000239-182128 & 00:02:39.71 & -18:21:28.8 & 29.83 & 0.28 & 0.16 & wish \\
  0000.1+4452 & J000244+450928 & 00:02:44.17 & +45:09:28.5 & 4.39 & 0.07 & 0.08 & wenss \\
  0000.1+4628 & J000242+464509 & 00:02:42.71 & +46:45:09.0 & 10.26 & 0.0 & 0.09 & wenss \\
  0000.2-2131 & J000249-211419 & 00:02:49.81 & -21:14:19.3 & 4.43 & -0.15 & 0.03 & wish \\
  0000.2-2251 & J000250-223437 & 00:02:50.77 & -22:34:37.7 & 9.52 & 0.29 & 0.14 & wish \\
  0000.3+2926 & J000252+294253 & 00:02:52.36 & +29:42:53.2 & 3.15 & 0.22 & 0.06 & wenss \\
\noalign{\smallskip}
\hline
\end{tabular}\\
Col. (1) WSRT name. \\
Col. (2) NVSS counterpart of the WSRT source. \\
Col. (3) R.A. from the NVSS catalog.\\
Col. (4) Dec. from the NVSS catalog.\\
Col. (5) Angular separation between the WSRT and the NVSS position: $R$.\\
Col. (6) Low frequency radio spectral index $\alpha_{low}$. \\
Col. (7) Uncertainty on the $\alpha_{low}$.\\
Col. (8) WSRT original survey: WENSS or WISH. 
\label{tab:LORCAT}
\end{table*}

\subsection{Flux density distributions at low-frequencies}
\label{sec:fluxes}
\begin{figure}[]  
           \includegraphics[height=9.5cm,width=6.8cm,angle=-90]{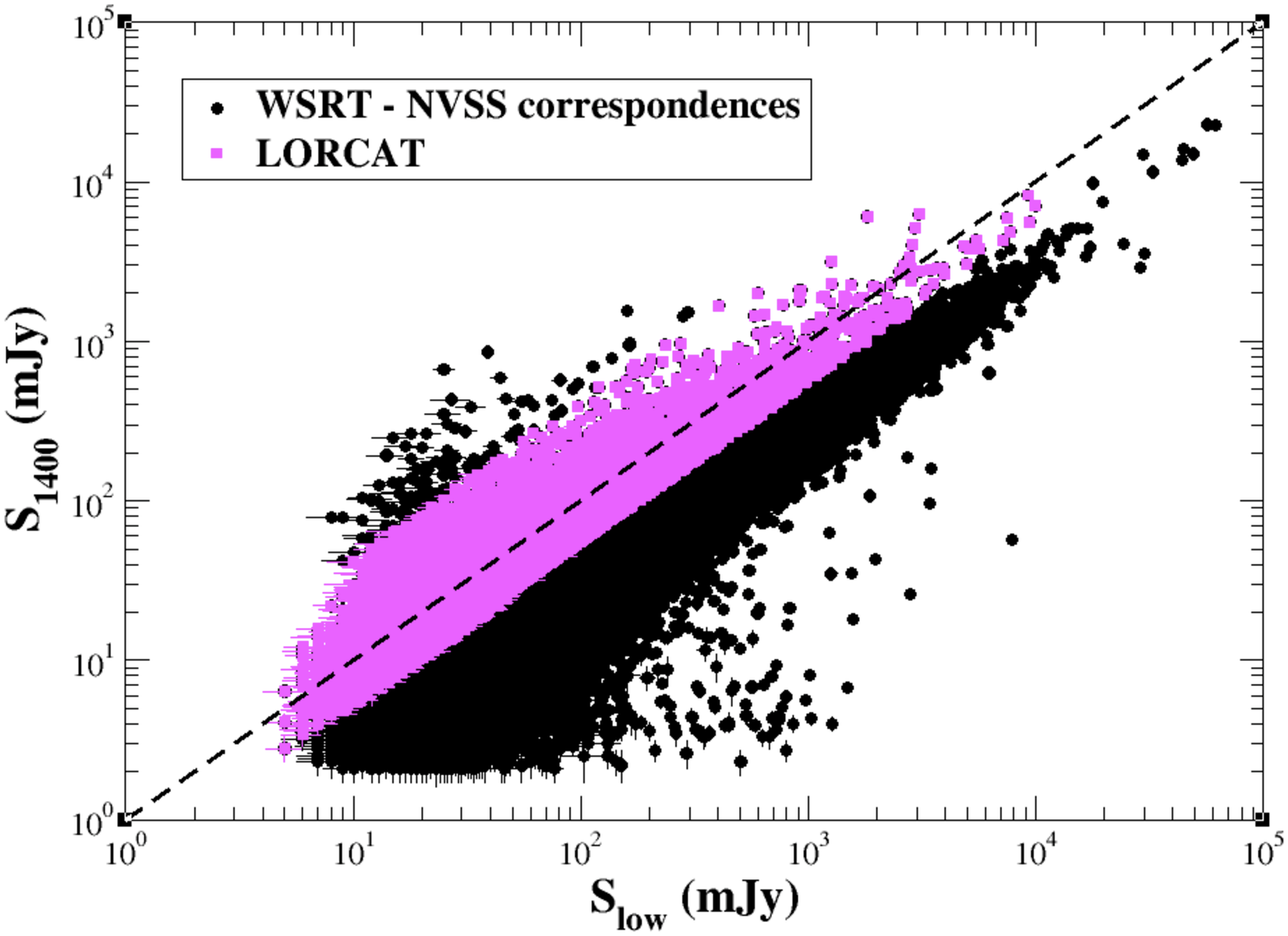}
           \caption{The flux density scatterplot. LORCAT sources (magenta circles) are shown in comparison 
                         with those associated in the whole WSRT-NVSS crossmatch (black circles). 
                        The dashed black line marks the radio spectral index $\alpha_{low}=$0.}
          \label{fig:fluxes}
          \end{figure}

\begin{figure*}[!ht]  
           \includegraphics[height=9.5cm,width=6.8cm,angle=-90]{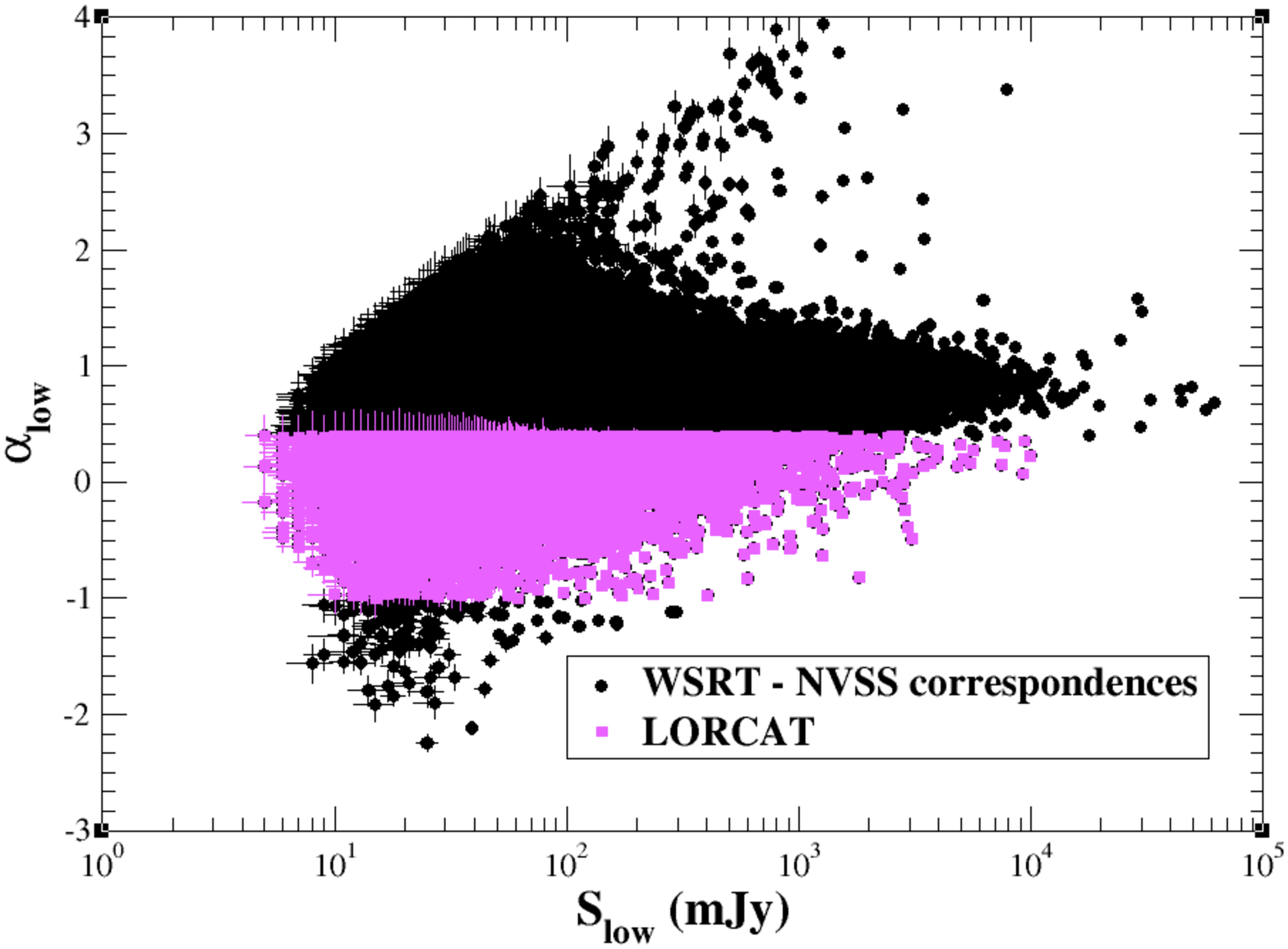}
           \includegraphics[height=9.5cm,width=6.8cm,angle=-90]{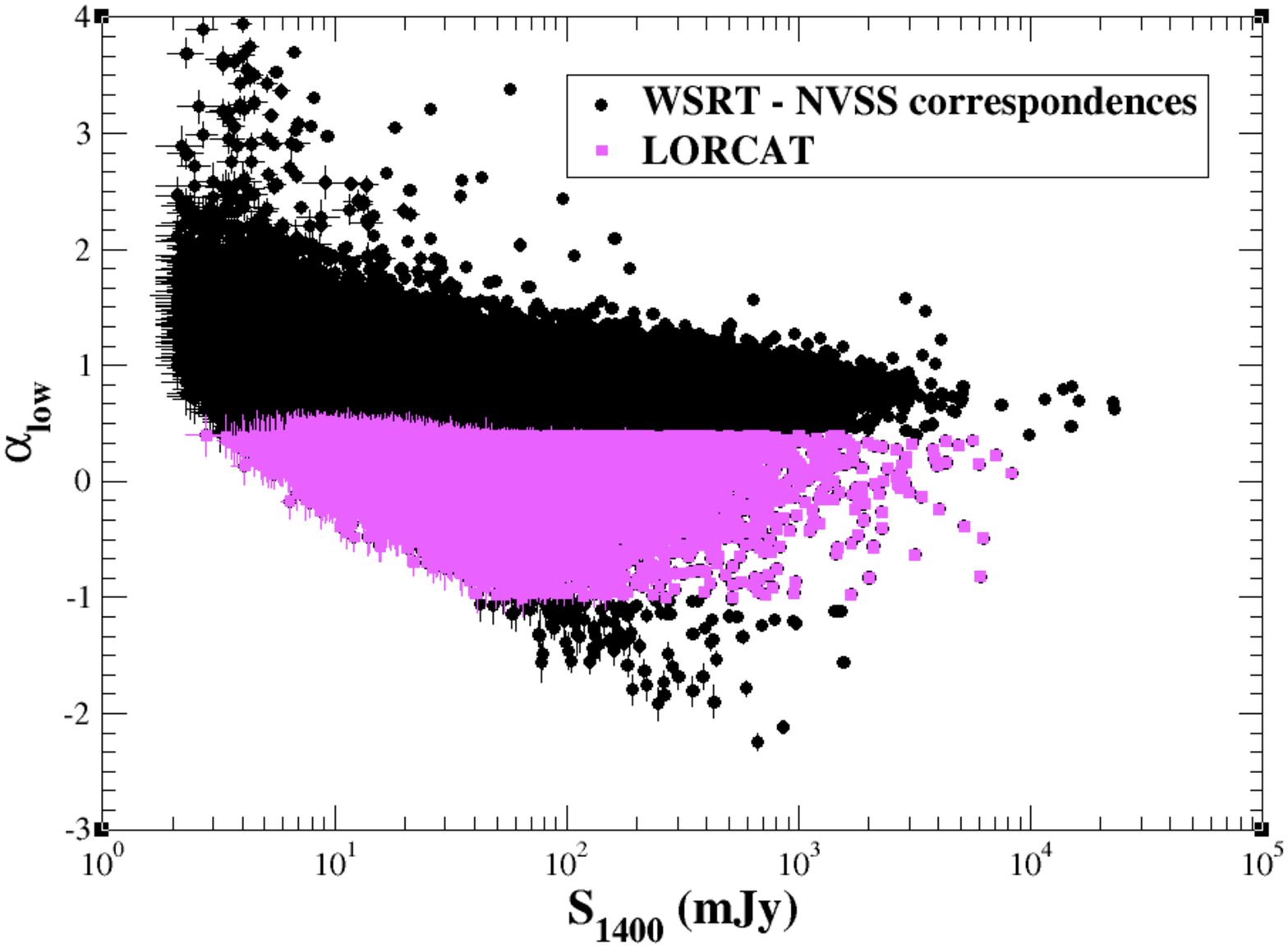}
           \caption{{\it Left panel}) The scatterplot of the low-frequency spectral index $\alpha_{low}$ with respect to the WSRT flux density.
                         {\it Right panel}) The same scatter plot where the NVSS flux density is reported on the x axis.}
          \label{fig:alfaflux}
          \end{figure*}

\begin{figure*}[]  
           \includegraphics[height=9.5cm,width=6.8cm,angle=-90]{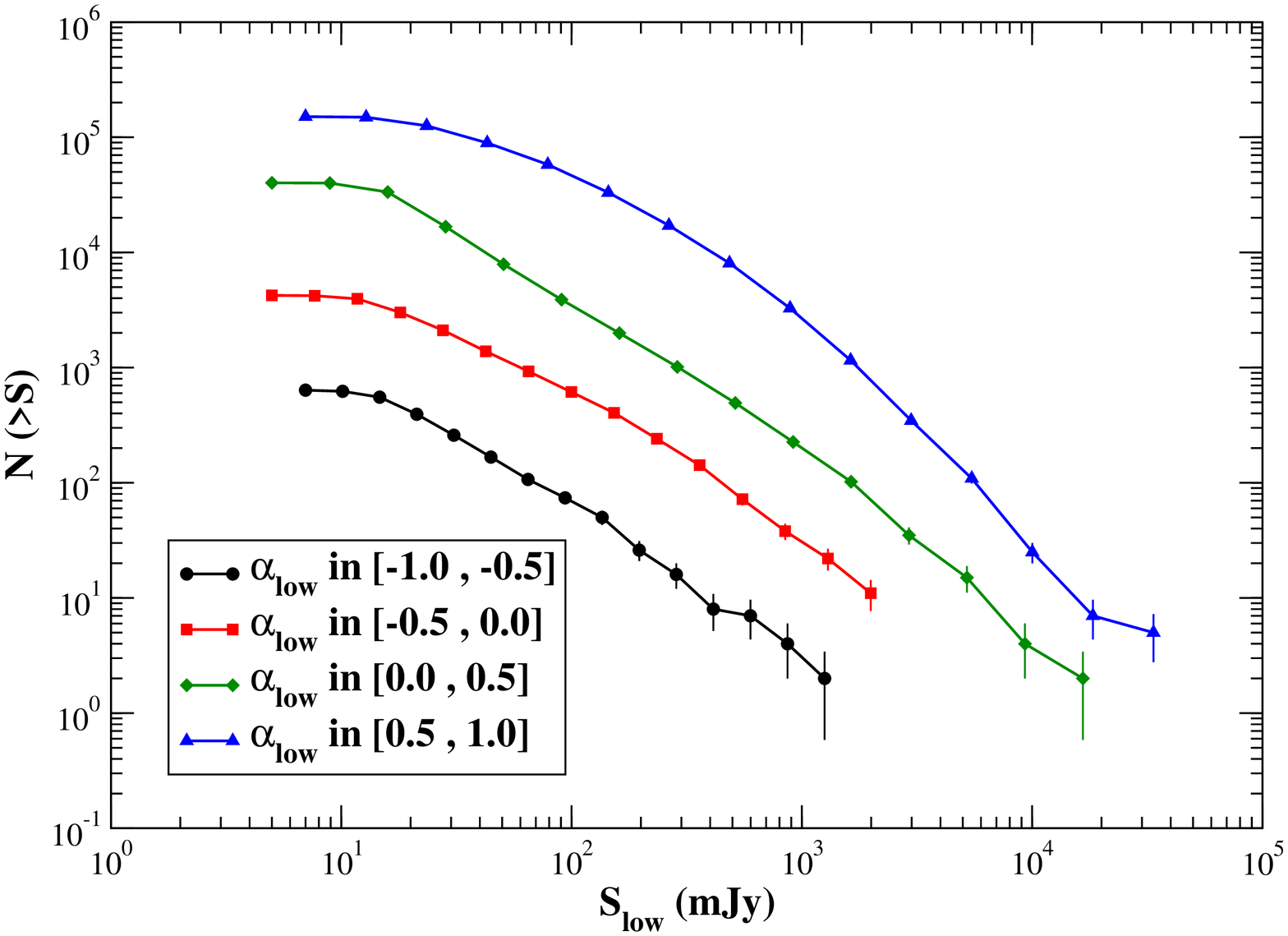}
           \includegraphics[height=9.5cm,width=6.8cm,angle=-90]{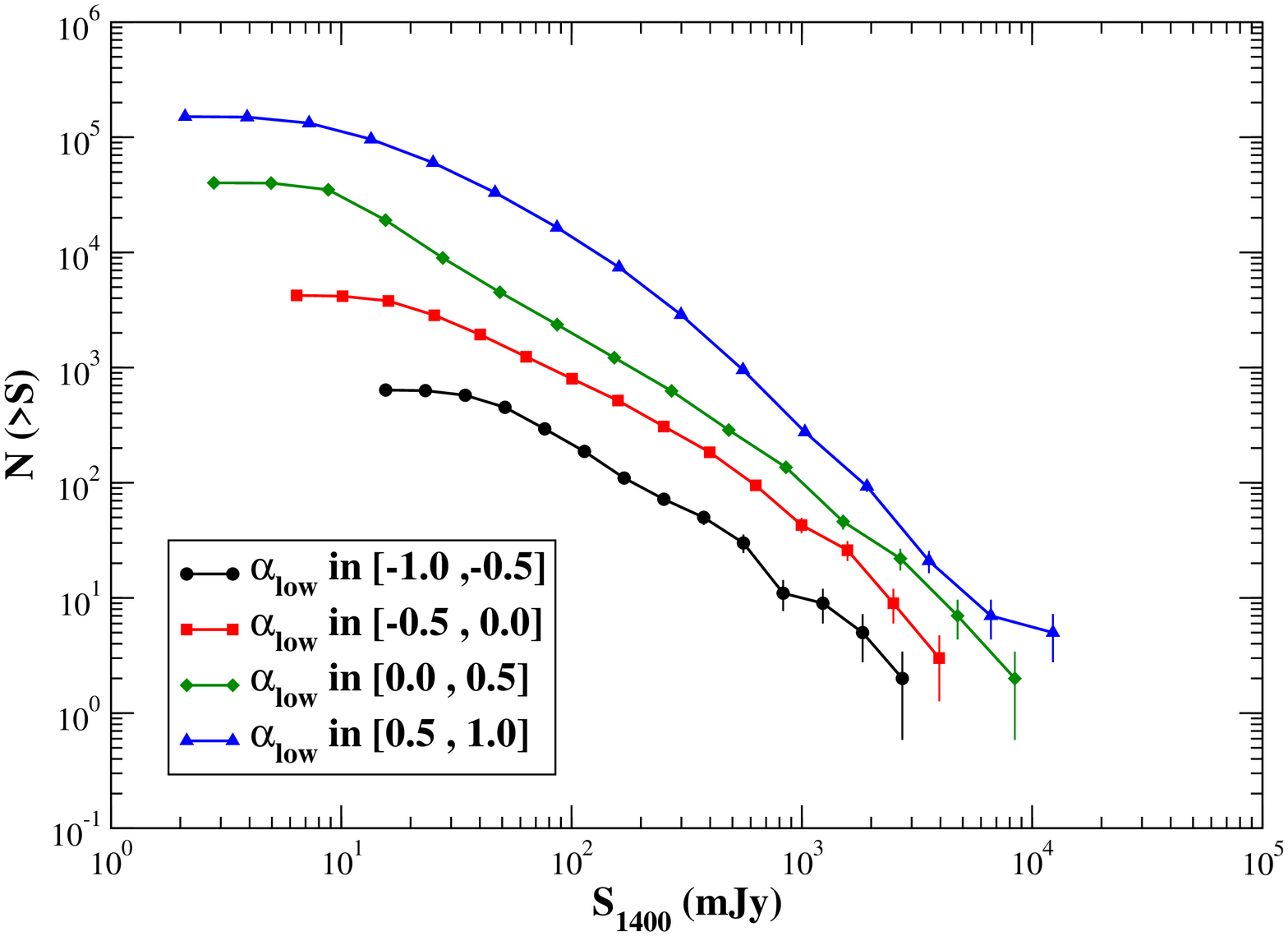}
           \caption{{\it Left panel}) The logN-logS distributions
                         evaluated for all the WSRT-NVSS correspondences with the low-frequency flux density $S_{low}$
                         for different range of $\alpha_{low}$ between the values -1 and 1.
                         {\it Right panel}) Same as the left panel for the logN-logS distributions computed using the NVSS flux densities at 1.4 GHz.}
          \label{fig:evol}
          \end{figure*}

\begin{figure}[]  
           \includegraphics[height=9.5cm,width=6.4cm,angle=-90]{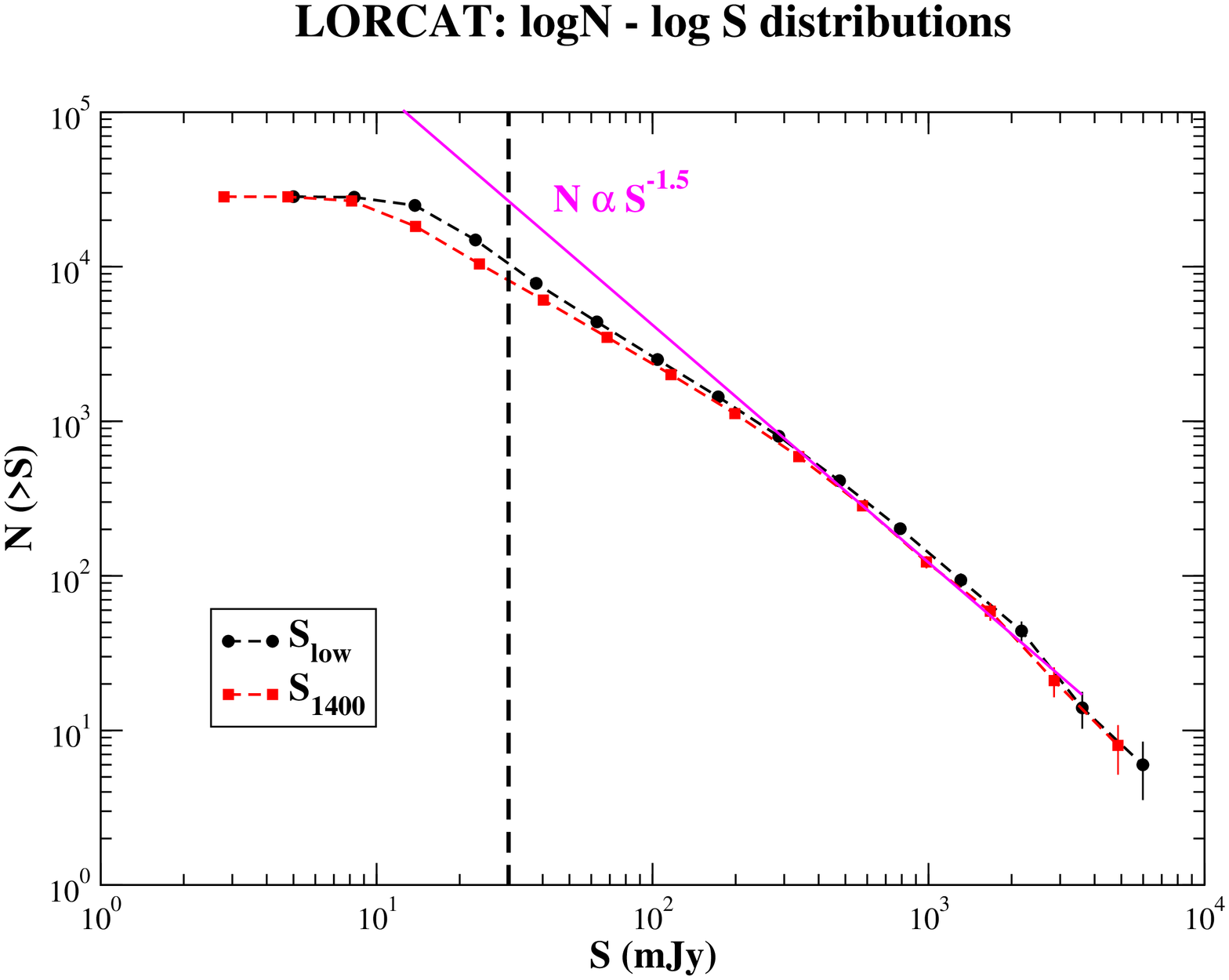}
           \caption{The logN-logS distributions of the LORCAT sample calculated with the WSRT flux density $S_{low}$ (black circles) and with 
                         that of the NVSS at 1.4 GHz (red squares).
                         A similar shape for the two logN-logS distributions of the LORCAT sample is expected since   
                        the $S_{low}$ and $S_{1400}$ flux densities are correlated (see Figure~\ref{fig:fluxes}).
                        The magenta line indicates the $N\propto\,S^{-1.5}$ relation expected from a uniform source distribution, while the vertical dashed black line
                        marks the completeness limit of the WSRT survey at 30 mJy.}
          \label{fig:logNlogS}
          \end{figure}

Comparing the radio flux densities at 1.4 GHz $S_{1400}$ and at low frequencies $S_{low}$ 
(i.e., 325 for the WENSS and 352 MHz for the WISH),
as shown in Figure~\ref{fig:fluxes}, there is a good match between the two WSRT and NVSS 
observations: bright sources below $\sim$1 GHz tend to be among 
the brightest also above 1 GHz. In Figure~\ref{fig:fluxes} we also report the line 
corresponding to a radio spectrum of $\alpha_{low}$=0.
Then in Figure~\ref{fig:alfaflux}, we also compare the low-frequency radio spectral index $\alpha_{low}$ 
with the archival WSRT and NVSS flux densities.
The logN-logS distribution for all of the WSRT-NVSS associations per range of low-frequency spectral indices between -1 and 1.5 
as well as that of our LORCAT are reported in Figure~\ref{fig:evol}. Then Figure~\ref{fig:logNlogS} shows the logN-logS distributions 
for the LORCAT sample for both the WSRT and the NVSS flux densities. 
These logN-logS distributions are in agreement with the evolution of the radio source counts \citep[e.g.,][]{condon84b,condon98}. 
These logN-logS distributions computed with both the $S_{low}$ and $S_{1400}$ 
for all the LORCAT sources appear to have the same shape. This is expected  
as the flux densities are mildly correlated, as shown in Figure~\ref{fig:fluxes}.
In Figure~\ref{fig:logNlogS} we also show the $N\propto\,S^{-1.5}$ line expected in the case 
of a uniform source distribution at not too large redshift (i.e. in a Euclidean universe). 
It is well know that blazars show a broken luminosity function due to the relativistic effects of their beamed emission \citep[e.g.][]{urry84},
this could be also reflected in the logN-logS distribution in agreement with that of the LORCAT.
However to prove this effect redshift estimates will be necessary for these low-frequency sources with flat radio spectra.

\section{Comparison with the Green Bank 6-cm radio source catalog}
\label{sec:cross}
A detailed identification of the complete LORCAT sample is out of the scope of the present analysis
and in particular a multifrequency analysis of the optical and the IR counterparts of the LORCAT sources
will be presented in a separate, forthcoming paper \citep{massaro14}.
However to understand the nature of the selected low-frequency flat spectrum radio sources
we performed a crossmatch with the Green Bank 6-cm radio source catalog 
(GB6)\footnote{http://heasarc.gsfc.nasa.gov/W3Browse/all/gb6.html} \citep{gregory96}
to investigate the spectral properties of the LORCAT sources at $\sim$5 GHz.

It is worth noting that among all the radio surveys at frequency grater than $\sim$1 GHz
the GB6 is the most recent one covering the largest portion of the LORCAT footprint since it was performed between 0\degr\ and +75\degr\
in declination. The GB6 radio source catalog is also complete above 50 mJy \citep{gregory96}.
Since the CRATES catalog have been compiled using the GB6 in the above range of declination,
a comparison with it is nested within the following analysis.

The total number of LORCAT sources within the GB6 footprint is 15814.
Assuming the difference $\Delta\alpha$ (i.e., $\Delta\alpha = \alpha_{high} - \alpha_{low}$)
between the low (i.e., $\alpha_{low}$) and the high-frequency (i.e., $\alpha_{high}$) spectral indices equal to zero 
with $\alpha_{high}$ defined as -1.85$\cdot log (S_{4850}/S_{1400})$, we computed the extrapolated flux density at 4.85 GHz
$S_{ex,4850}$ for the LORCAT sources to determine those expected to be detected in the GB6.
We found that above the completeness threshold of the GB6, there are 3219 LORCAT sources having $S_{ex,4850}>$50mJy. 

\begin{figure}[]  
           \includegraphics[height=9.5cm,width=6.2cm,angle=-90]{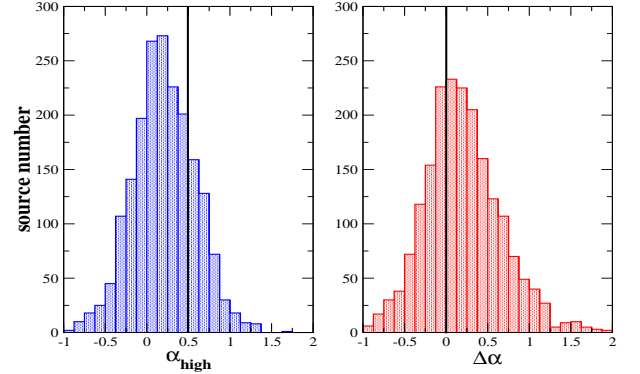}
           \caption{Left panel) The distribution of the high-frequency spectral index $\alpha_{high}$
                        computed for all the LORCAT sources with a counterpart in the GB6 survey within 
                        their radio positional uncertainties (see Section~\ref{sec:cross} for more details).
                        It is worth noting that a significant fraction of sources having flat spectra at low 
                        frequencies between $\sim$300 MHz and $\sim$1 GHz (i.e., $-1<\alpha_{low}<$0.4)
                        appear to be relatively flat, according to the general definition (i.e., $\alpha_{high}<$0.5 marked by the vertical black line), 
                        also at high frequencies between $\sim$1 GHz and $\sim$5 GHz.
                        Right panel) The distribution of the $\Delta\alpha = \alpha_{high} - \alpha_{low}$ for the LORCAT-GB6 radio correspondences.
                        Radio sources with flatter high-frequency spectrum with respect to the low-energy one have $\Delta\alpha<0$ 
                        while those steepening at high frequencies show $\Delta\alpha>0$. The vertical dashed line marks the threshold $\Delta\alpha$=0.}
          \label{fig:alfahigh}
          \end{figure}

Then searching the correspondences between the LORCAT and the GB6 catalogs,
we found 1942 out of the 3219 (i.e., $\sim$60\%) are detected at 6-cm 
within their positional uncertainty regions at 1$\sigma$ level of confidence,
computed between the NVSS and the GB6 positions.
In particular only 875 out of these 1942 radio sources show a flux density $S_{4850}$ 
above the completeness limit of the GB6 survey.
The distribution of the high-frequency spectral index $\alpha_{high}$ computed with the observed 
$S_{4850}$ in the GB6 for the 2834 LORCAT-GB6 associations
is reported in Figure~\ref{fig:alfahigh} together with their $\Delta\alpha$ histogram.
More than $\sim$75\% of the 1942 LORCAT sources detected in the GB6 
still have a ``flat'' radio spectrum at frequencies above $\sim$1 GHz (see Figure~\ref{fig:alfahigh}),
according to the canonical, widely accepted definition of flat spectrum radio sources 
(i.e., $\alpha_{high}<$0.5) \citep[e.g.,][and references therein]{kellermann74,condon88}.
This strongly support our definition of low-frequency ``flat'' radio spectra (see Section~\ref{sec:index}).
However, a significant fraction (i.e., $\sim$60\%)of these GB6-LORCAT sources appear to have radio spectra 
that steepens toward higher frequencies (i.e., $\Delta\alpha>0$).

\begin{figure}[]  
           \includegraphics[height=9.5cm,width=6.cm,angle=-90]{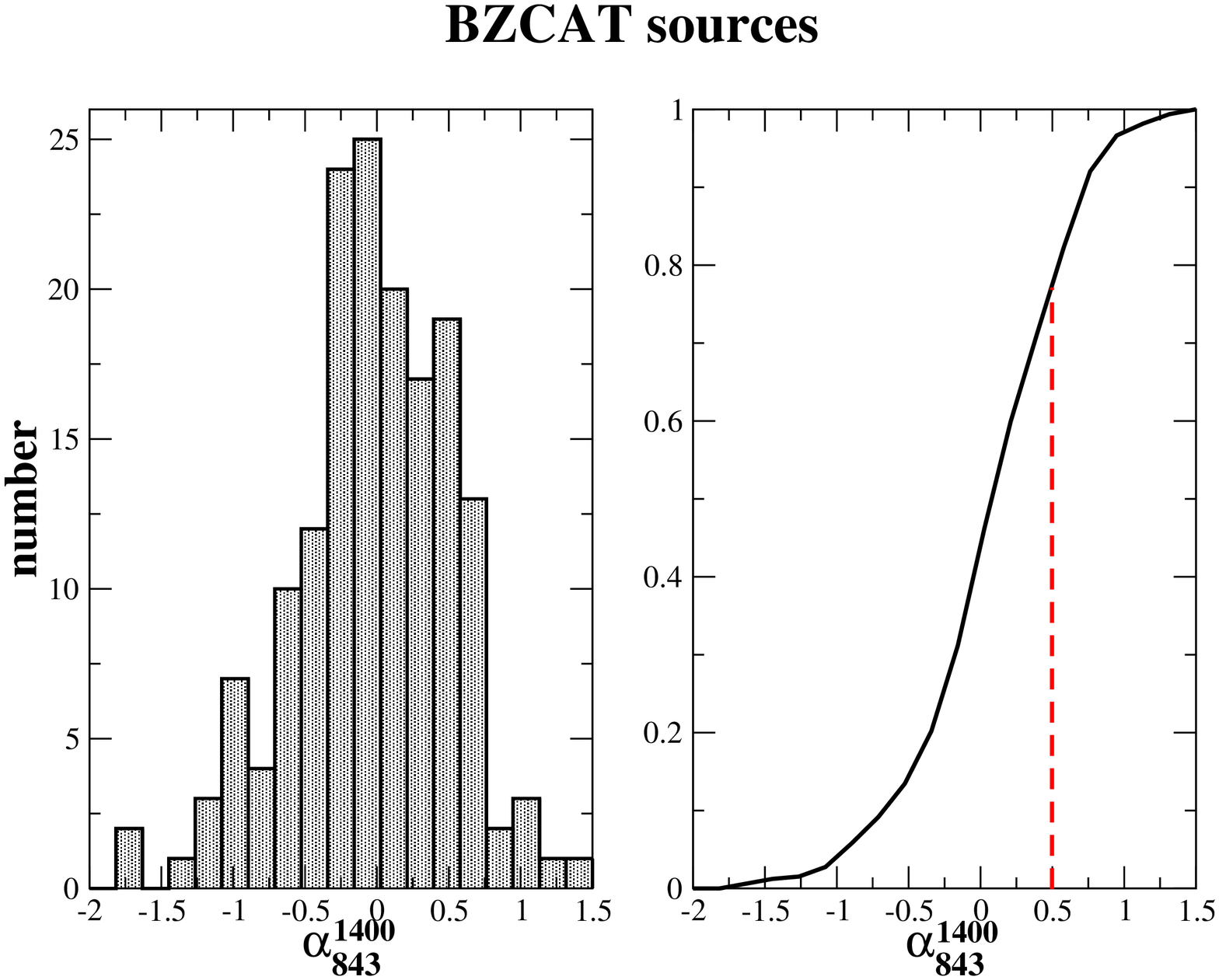}
           \caption{The distribution of the radio spectral index $\alpha^{1400}_{843}$ for all the known blazars that lie
                          within the NVSS and in the SUMSS footprints (left panel).
                          The cumulative distribution is shown on the right panel where the red dashed line marks the $\alpha^{1400}_{843}=$0.5,
                          according to the canonical definition of flat spectrum radio sources.}
          \label{fig:sumss_nvss}
          \end{figure}

{ In particular, it is worth mentioning that the subclass of BZQs generally show flat high-frequency spectra, 
thus LORCAT sources with steep high-frequency spectra may not actually be BZQs \citep[e.g.,][]{condon83}.
However, $\Delta\alpha>0$ is occurring for a small fraction (i.e., $\sim$5\%) of the known blazars listed in the \bzcat
and these are all classified as BL Lac objects.}
In Figure~\ref{fig:sumss_nvss} we also report the radio spectral index $\alpha^{1400}_{843}$ evaluated for all the \bzcat\ blazars 
that have radio observations in the NVSS and in the Sydney University Molonglo Sky Survey \citep[SUMSS;][]{mauch03} at 843 MHz.
It is evident that blazars show a clear steepening a higher frequencies in agreement with that found for the LORCAT sources.
In addition, there could also be the possibility that these radio spectra are intrinsically mildly curved \citep[e.g.,][]{howard65,kellermann69,pauliny72}.
It is known that spectral curvature appears at higher frequencies in the sub millimeter data \citep[e.g.,][]{giommi07,giommi12}.

Finally, we note the presence of radio sources with $\Delta\alpha<0$ in the LORCAT sources
might indicate that the low-frequency emission could be contaminated by that of 
extended components. These cannot be resolved with the large beam of the low-frequency survey and in general present steep spectra 
\citep[see e.g.,][for a recent discussion]{massaro13b}.

\section{Summary and conclusions}
\label{sec:conclusions}
We have assembled a low-frequency radio catalog of flat spectrum sources (LORCAT)
built by combining the radio observations of the two main WSRT surveys (i.e., WENSS and WISH) at 325 MHz and 352 MHz, respectively,
with those of the NVSS at 1.4 GHz.
The main goals underlying the creation of this catalog are similar to those of the CRATES \citep{healey07} since both 
can be used in the future to search for new, unknown blazar-like counterparts of the $\gamma$-ray sources 
\footnote{The LORCAT catalog has been already used for the $\gamma$-ray source associations 
that will be released with the the next \fer\ catalog actually in preparation.}.

We defined a new criterion to associate WSRT and NVSS sources improving our previous analyses \citep{ugs3,ugs6} and 
we provided a new definition of flat spectrum radio sources at low-frequencies based on the distribution of the spectral index $\alpha_{low}$
between 325 MHz and 1.4 GHz found for blazars in the \bzcat.
Sources with radio analysis flags as well as double matches between the radio surveys have been excluded from our final list.
Thus the LORCAT sample comprises 28358 radio sources including $\sim$667 known blazars having $-1<\alpha_{low}<$0.4.

Then we also compared our LORCAT catalog with the the Green Bank 6-cm (GB6) radio catalog, 
since it is the most recent radio survey covering the largest fraction of the LORCAT footprint at higher frequency (i.e., $\sim$5 GHz).
We found that a significant fraction of the LORCAT sources with extrapolated flux densities at $\sim$5 GHz
above the completeness threshold of the GB6 are detected (i.e., $\sim$86\%).
In addition they also appear to be ``flat'' spectrum radio sources above $\sim$1 GHz, according to the canonical definition 
(i.e., $\alpha_{high}<$0.5) \citep[e.g.,][and references therein]{condon88}.
The lack of detections for a small fraction of the LORCAT sources in the GB6 footprint could be explained in terms of 
a spectral steeping toward high frequencies (i.e., a mild curvature)  as already observed in blazars.

Finally, we highlight that to investigate the nature of the LORCAT sources, aiming to identify 
the fraction of $\gamma$-ray blazar candidates associable to \fer\ sources,
a detailed analysis of the IR and optical properties is necessary. 
It will be presented in a separate, forthcoming paper \citep{massaro14}.

\acknowledgements
We are in debt with our anonymous referee for many helpful comments and for all the checks performed on our tables.
This work is supported by the NASA grants NNX12AO97G and NNX13AP20G.
R. D'Abrusco gratefully acknowledges the financial 
support of the US Virtual Astronomical Observatory, which is sponsored by the
National Science Foundation and the National Aeronautics and Space Administration.
The work by G. Tosti is supported by the ASI/INAF contract I/005/12/0.
P.S.C. is grateful for support from the NSF through the NSF Graduate Research Fellowships Program Grant DGE1144152
The WENSS project was a collaboration between the Netherlands Foundation 
for Research in Astronomy and the Leiden Observatory. 
We acknowledge the WENSS team consisted of Ger de Bruyn, Yuan Tang, 
Roeland Rengelink, George Miley, Huub Rottgering, Malcolm Bremer, 
Martin Bremer, Wim Brouw, Ernst Raimond and David Fullagar 
for the extensive work aimed at producing the WENSS catalog.
Part of this work is based on archival data, software or on-line services provided by the ASI Science Data Center.
This research has made use of data obtained from the high-energy Astrophysics Science Archive
Research Center (HEASARC) provided by NASA's Goddard
Space Flight Center; the SIMBAD database operated at CDS,
Strasbourg, France; the NASA/IPAC Extragalactic Database
(NED) operated by the Jet Propulsion Laboratory, California
Institute of Technology, under contract with the National Aeronautics and Space Administration.
Part of this work is based on the NVSS (NRAO VLA Sky Survey):
The National Radio Astronomy Observatory is operated by Associated Universities,
Inc., under contract with the National Science Foundation. 
TOPCAT\footnote{\underline{http://www.star.bris.ac.uk/$\sim$mbt/topcat/}} 
\citep{taylor05} for the preparation and manipulation of the tabular data and the images.

{Facilities:}\facility{WSRT}, \facility{VLA}, \facility{GBT}.

{}

\end{document}